\documentstyle[prl,aps,multicol]{revtex}

\input BoxedEPS
\SetepsfEPSFSpecial
\HideDisplacementBoxes

\begin{document}

\title{Ferromagnetic resonance in periodic particle arrays}
\author{S.\ Jung,$^1$ B.\ Watkins,$^{1,2}$ L. DeLong,$^{1,2}$ J.\ B.\ Ketterson$^1$ and V.\ Chandrasekhar$^1$}
\address{ $^1$Department of Physics and Astronomy, Northwestern University, Evanston, IL 60208, USA \\ 
          $^2$Department of Physics and Astronomy, University of Kentucky, KY 40508, USA \\   }

\maketitle

\begin{abstract} We report measurements of the ferromagnetic resonance (FMR) spectra of arrays of submicron size
periodic particle arrays of permalloy produced by electron-beam lithography.  In contrast to plane ferromagnetic
films, the spectra of the arrays show a number of additional resonance peaks, whose position depends strongly on
the orientation of the external magnetic field and the interparticle interaction.  Time-dependent micromagnetic
simulations of the ac response show that these peaks are associated with coupled exchange and dipolar spin wave
modes.  
\end{abstract}


\begin{multicols}{2}
The physics of small ferromagnetic particles has become of increasing interest in recent years, driven to a great
extent by  their potential applications in high-density digital storage.  As the size of the particles decreases,
it is important to have access to experimental tools that can probe magnetic properties on the scale of
nanometers.  Conventional magnetic probes such as magnetometry are not sensitive enough to probe the properties
of individual nanometer-scale particles.  Recently developed tools such as magnetic force microscopy (MFM)
\cite{martin} can indeed image nanometer scale features, but their results are not always easy to interpret in
terms of the intrinsic parameters of the ferromagnet.  From this perspective, ferromagnetic resonance (FMR) is a
powerful tool to probe the fundamental magnetic properties of ferromagnetic particles \cite{heinrich}.  As has
been known for many years, FMR spectra are sensitive to the detailed geometry of the sample, and the FMR
resonance shape depends on the intrinsic dissipation mechanisms in the ferromagnet.  In addition, information
about magnetic excitations such as spin waves can also be obtained.  In spite of this, FMR has not been used
extensively to probe the properties of nanometer scale ferromagnets.  It is only recently that techniques such as
magnetic resonance force microscopy (MRFM) \cite{wago} have been applied to small magnetic structures, but such
investigations are still in their infancy.

Consider then a ferromagnetic particle in a uniform external magnetic field $H_0$.  Its magnetic moment $M$ will
evolve according to the Landau-Lifschitz equation \cite{aharoni}
\begin{equation}
\frac{d\mathbf{M}}{dt}=-\gamma \mathbf{M} \times \mathbf{H}_{eff} -\frac{\gamma \alpha}{M_s} \mathbf{M} \times
(\mathbf{M}
\times  \mathbf{H}_{eff}) 
\label{dynamic}
\end{equation}  where $\mathbf{M}$ is the magnetization, $\gamma=ge/2mc$ the gyromagnetic ratio ($g$ being the
Land\'e factor, and $m$ the mass of the electron),
$\mathbf{H}_{eff}$ the effective local field, $M_s$ the saturation magnetization, and $\alpha$ the damping
constant.  This causes a precession of the moment about the direction of the field at a frequency
$\omega_0$.  For an ellipsoidal ferromagnet body (with a uniform magnetization), $\omega_0$ is given by the
Kittel equation
\cite{kittel}
\begin{equation}
\omega_0^2=\gamma^2 [H_0+(N_x-N_z)M][H_0+(N_y-N_z)M]
\label{kittel}
\end{equation} where the $N_x$, $N_y$, and $N_z$ are the demagnetization factors along the $x$, $y$ and $z$
directions which satisfy the relation $N_x + N_y + N_z = 4 \pi$, and the $z$ direction is defined by the
orientation of the external field $H_0$ and the magnetization $M$.  
An ac magnetic field
$H_{ac}$ at this resonant frequency applied perpendicular to $H_0$ will couple to a uniform precession of $M$
about the direction of $H_0$, resulting in absorption of energy from the ac field.

In addition to the uniform precession mode, one may also have non-uniform, or spin-wave, modes of precession. 
Kittel \cite{kittel2} considered the case where the exchange interaction provided the dominant correction to
the mode frequency.  In his model, additional resonance peaks would be observed at frequencies $\omega_p= D k_p^2
+ \omega_0$, where $D$ is a constant dependent on the exchange interaction between neighboring spins, and $k_p$ is
the wave vector of the spin-wave which is quantized due to the pinning of the surface spins by surface
anisotropy.  For such exchange resonance modes, the resonance frequency is always larger than that of the
uniform mode, which implies that the resonance occurs at an applied field less than that of the uniform mode
\cite{seavey}.  The resonance frequencies depend on the size of the particle through the
quantization of the wave vector $k_p$.  In contrast, when the long-range dipolar interactions are stronger than
the exchange interactions, as is the case in larger particles, the resulting magnetostatic or Walker modes
\cite{walker} may be higher or lower in frequency than the uniform mode (indeed, the uniform mode itself can be
considered to be a magnetostatic mode), and their frequency is expected to be independent of the size of the particle.  For
particles of intermediate size, where the exchange and dipolar energies are comparable, one may have spin wave
modes whose dynamics are determined by both short range exchange interactions and long range dipolar fields.  In
this case, numerical simulations show that the resonance frequencies can be both above and below the frequency
of the uniform mode, and the resonance frequencies depend in a complicated manner on the size of the particle
\cite{voltairas,arias}.  To our knowledge, these coupled exchange modes have not been clearly observed in a FMR
measurement.  

In this Letter, we present FMR measurements on square periodic arrays of submicron size permalloy particles.  In
addition to the resonance peak expected for a uniform ferromagnetic film, we observe a number of sideband peaks
which we associate with spin wave modes in the particles.  The position of the peaks can be very sensitive to the
relative orientation of the applied external field with respect to the sample.  In addition, when the
interparticle spacing is very close, the position of the resonance fields show a four-fold symmetry as a function
of the relative orientation of the external field, reflecting the underlying symmetry of the periodic lattice. 
Numerical micromagnetic simulations of the FMR response show that these sideband peaks are associated with
coupled exchange and dipole spin wave modes.

\begin{figure}[h!]
\vspace{-7.0 cm}
\begin{center}
\BoxedEPSF{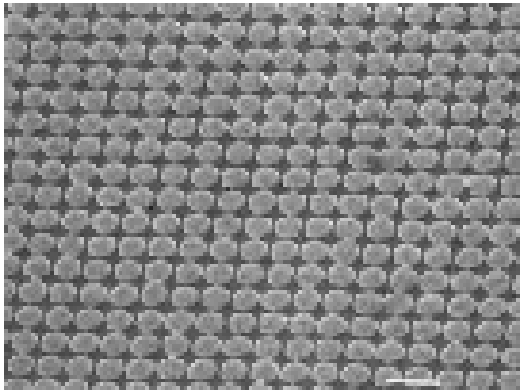 scaled 1000}
\end{center}
\vspace{-3.5 cm}
\caption{Scanning electron micrograph a square array of circular permalloy disk of diameter 0.5 $\mu$m, spaced a distance 
0.6 $\mu$m apart.  The size bar corresponds to 1 $\mu$m.}
\label{figure1}
\end{figure}

\vspace{-1.5 cm}      
The ferromagnetic particle arrays in this study were fabricated by conventional electron beam lithography on
thermally oxidized Si substrates.  After patterning, the permalloy (Ni$_{79}$Fe$_{21}$) films were electron-beam
evaporated, resulting in polycrystalline films of thickness $t$ ranging from 70 nm to 90 nm.  The arrays
have circular ferromagnetic elements, with diameters $d$ ranging from 100 nm to 500 nm. 
The square lattice constant
$a$ was varied from 150 nm to 1 $\mu$m, and the total area of each array was 400 $\mu$m $\times$ 400 $\mu$m. 
Over 40 different arrays were fabricated and measured; only a few representative samples are discussed here. 
Figure 1 shows a scanning electron micrograph (SEM) of one of the particle arrays.    The samples were
measured at room temperature in a Varian electron spin resonance cavity spectrometer with an operating frequency
of 9.37 GHz, and the orientation of the static dc field could be varied with respect to the sample
substrate. 

\begin{figure}[t]
\BoxedEPSF{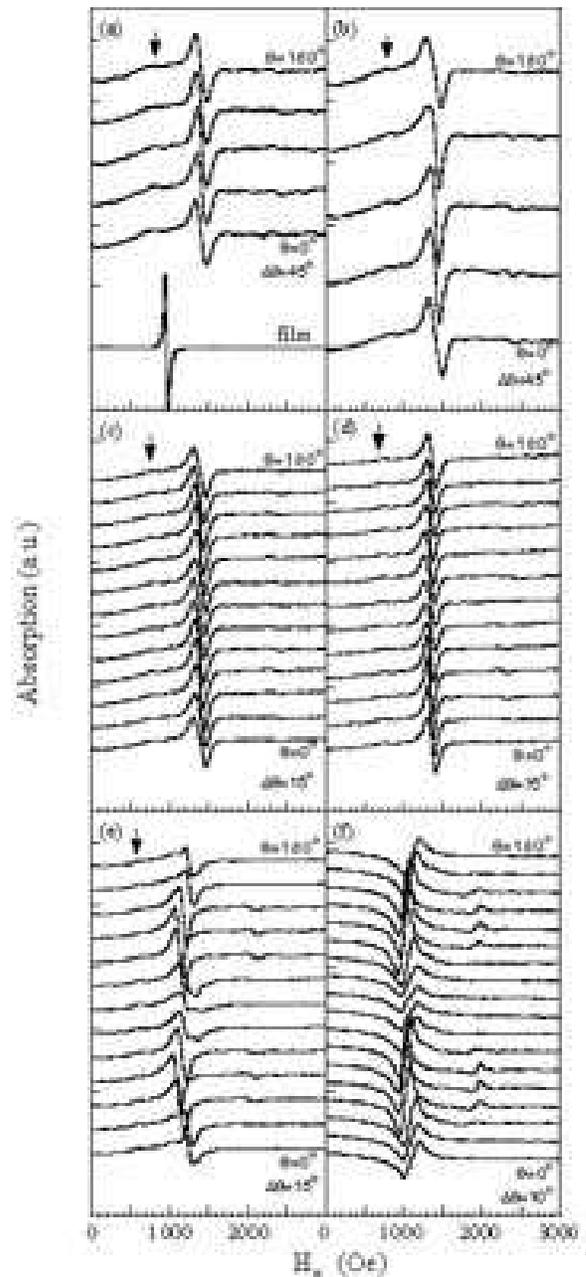 scaled 1000}
\caption{(a)-(e) Room temperature derivative FMR spectra for square lattices of circular elements of diameter $d=$ 0.5 $\mu$m with spacing $a=$ (a) 1.5, (b) 1.1, (c) 0.9,
(d) 0.8, (e) 0.6 $\mu$m.  (f) Similar data for a lattice with $d=0.53$ $\mu$m and $a=$0.6 $\mu$m.}
\label{figure2}
\end{figure}

Figure 2(a) shows derivative FMR spectra for a square array of circular permalloy particles with $d=0.5$ $\mu$m,
$t=85$ nm, and $a=1.5$ $\mu$m, as a function of the angle of the magnetic field with respect to one axis of the
square array (the magnetic field is always in the plane of the substrate).  A large resonance peak is observed at
$\sim$1400 Oe, which corresponds to the uniform resonance mode.  For reference, we also show the equivalent
spectrum for a plane permalloy film of thickness 60 nm, with $H_0$ applied in the plane of the film.  The peak
for the particle array occurs at a slightly higher magnetic field in comparison to that of the plane
ferromagnetic film due to demagnetization effects.  As in the plane film, the position of the peak does not shift
as a function of the angle of the magnetic field in the substrate plane, which might be expected from the
circular symmetry of the array elements.  In addition to the peak for the uniform mode, however, the circle array
also shows a small peak at lower field (marked by an arrow).  As we noted above, similar peaks have been
observed in uniform ferromagnetic films and small ferromagnetic particles, where they are associated with
standing exchange spin wave resonances in the sample \cite{seavey,toneguzzo}.  In our case, however, they are
associated with exchange spin-wave modes with a small contribution from dipolar interactions, as can be seen by
noting their evolution as the distance between the circular particles is reduced.  Figures 2(b)-(e) show
equivalent FMR spectra for circle array samples evaporated at the same time as the sample of Fig. 2(a).  The
nominal diameters for the circles (0.5
$\mu$m) are the same; the only parameter that is changing is $a$, which varies from
$a=1.5$ $\mu$m to 600 nm.  As the interparticle spacing is reduced, the position of the peak denoted by the
arrows move down in field slightly.  Since the only interaction that could be changing as the
spacing is reduced is the dipolar coupling between particles, this indicates that the resonance frequency of
these modes also depends (albeit weakly) on dipolar interactions in the system.

\begin{figure}
\vspace{-0.25 cm}
\BoxedEPSF{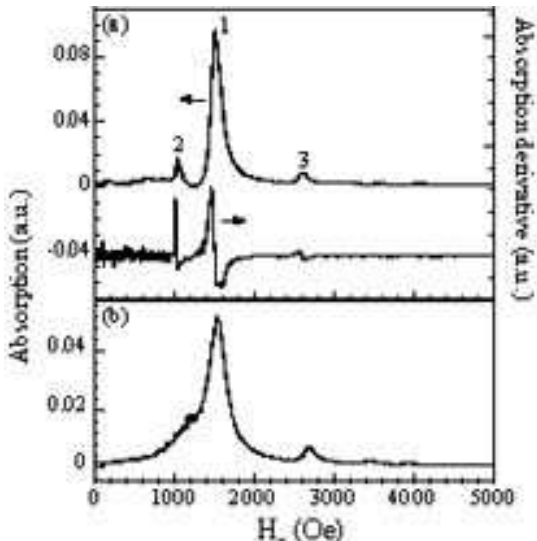 scaled 950}
\caption{(a)  Calculated absorption spectrum for a $d=$0.5 $\mu$m particle with $M_s=8.47 \times 10^3$ A/m, exchange stiffness $A=1.3 
\times 10^{-13}$ J/m and damping constant $\alpha=$0.05;  (b)  Identical to (a), except with $A=0$.}
\label{figure3}
\end{figure}

Three additional observations indicate that magnetostatic interactions between the particles strongly influence
the FMR spectrum of the arrays.  First, at small separations, a number of new resonance peaks are observed.  This
is most clearly seen in Fig. 2(e), both near and below the uniform mode resonance, as well as at higher magnetic
fields (e.g., at $\sim$ 2100 Oe).  Second, the amplitude of these additional peaks varies as a function of the
angle of the magnetic field in the plane of the film.  For example, looking at Fig. 2(e) again, the resonance at
$\sim$ 2100 Oe is most clearly developed when the external magnetic field is oriented along a diagonal of the
square lattice.  Figure 2(f) shows a similar but much more pronounced effect in another circle array sample
fabricated in a different run.  Third, the field at which the uniform resonance mode occurs oscillates as a
function of the angle of the applied magnetic field, as seen most clearly in Fig. 2(e) and (f).  This indicates
that the dipolar interactions between particles also influence the energy of the uniform mode. 

In order to understand the nature of the resonance modes, we have performed time-dependent micromagnetic
calculations of the ac response of isolated ferromagnetic particles by numerically solving Eq.(\ref{dynamic}),
using the public OOMMF micromagnetic code solver \cite{oommf}.  The FMR absorption spectrum is obtained by
applying a small ($\sim$ 10 Oe) transverse ac field at a frequency of 9.37 GHz at each value of the dc magnetic
field, and calculating the amplitude of the steady-state magnetization response at this frequency.  The details
of the calculation will be discussed elsewhere \cite{jung}.  Figure 3(a) shows the resulting absorption spectrum
for a $d=$0.5 $\mu$m permalloy particle with saturation magnetization $M_s=8.47
\times 10^3$ A/m, exchange stiffness $A=1.3 \times 10^{-13}$J/m and damping constant $\alpha=0.05$ (the
derivative of this curve is also shown to facilitate comparison to the experimental curves in Fig. 2).  In
addition to the large peak corresponding to the uniform mode, satellite peaks are also observed on
both sides of the uniform mode peak, similar to what is observed experimentally.  The precise positions of these
peaks do not match the experimental data, since they are very sensitive to the exact values of the parameters
used in the calculations.  However, the qualitative trend is very similar.  In order to demonstrate that the
resonance peaks correspond to coupled exchange-dipolar spin-wave modes, we show in Fig. 3(b) the results of
another calculation in which all parameters are identical to the calculation of Fig. 3(a), except that the
exchange stiffness $A$ is set to zero, effectively eliminating any exchange contribution to the spin-waves. 
Although the position of the peak corresponding to the uniform mode is not affected, the positions of the peaks
corresponding to the spin-wave modes change.  In fact, the mode corresponding to the peak
below the uniform mode peak essentially disappears, showing that this peak is associated primarily with an
exchange spin-wave mode.  The resonance fields of the modes above the uniform mode are shifted slightly.  With
$A=0$, these spin-wave modes correspond to pure magnetostatic or Walker modes; the difference in the position of
these resonances compared to those of Fig. 3(a) show that the resonances of Fig. 3(a) involve a coupling of the 
exchange and dipolar contributions to the spin-wave modes.

\begin{figure}
\vspace{-0.25 cm}
\begin{center}
\BoxedEPSF{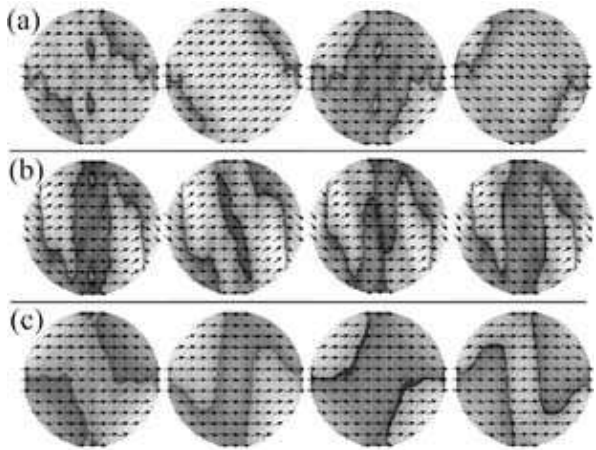 scaled 1000}
\end{center}
\caption{Each panel represents the time evolution of the magnetization distribution through one cycle of the ac field 
(at phases $\sim$0, $\pi/2$, $\pi$, $3\pi/2$) for a $d=0.5$ $\mu$m
particle at magnetic fields corresponding to the peaks labeled 1-3 in Fig. 3(a). (a) peak 1; (b) peak 2; (c) peak 3.
$H_0$ is aligned along the $x$-axis, and the grayscale plot denotes the angle between the $y$ component of $M$ and
the $x$-axis.}
\label{figure4}
\end{figure}

Further evidence of the nature of the modes corresponding to the resonance peaks observed in Fig. 3(a) can be
obtained by visualizing the time-dependent magnetization at the corresponding values of magnetic field.  The four
panels in Fig. 4(a) show the magnetization distribution during one period of the ac field at the dc magnetic
field corresponding to the uniform mode, denoted by the arrow labelled `1' in Fig. 3(a).  In order to make the
picture clearer, each arrow in a panel denotes the magnetization over an area much larger than the cell
size used in the calculation (5 nm $\times$ 5 nm).  The uniform nature of the mode can be clearly seen in the
fact that the motion of each arrow is almost identical to that of its neighbors, except for the areas around the
edge of the circle.  Figure 4(b) shows a similar evolution for the resonance peak at $\sim$1000 Oe (labeled `2'
in Fig. 3(a)), below the uniform mode peak.  In this case, the ac response of the system consists of small
oscillations about a very non-uniform static magnetization distribution, or in other words, a spin-wave mode. 
The fact that this mode almost completely disappears when the exchange stiffness $A=0$ suggests that the major
contribution to the energy of this mode comes from the standing exchange spin-wave, as discussed by Kittel. 
Figure 4(c) shows the time evolution over one period of the ac field for the resonance peak at 2500 Oe, labeled
`3' in Fig. 3(a).  As can be seen, the nature of this mode is quite different from either the uniform mode or the
primarily spin-exchange mode we have just discussed.  In this case, the magnetization in the center of the
particle is mostly static and aligned along the direction of the magnetic field; the response of the particle is
confined primarily to small oscillations of the magnetization near the edges.  Due to shape demagnetization effects, the
magnetization near the edges is non-uniform, and will give rise to an exchange contribution to the mode energy. 
We believe it is also these edge effects which give rise to the small dipolar contribution to the spin-wave mode
labeled `1.'  The confinement of the non-uniform magnetization to the edges of the circle also suggests why these
modes may be influenced strongly by the dipolar field arising from another ferromagnetic particle placed in close
proximity, as we have seen in our experimental results.  In principle, one could model an array of circular
particles, but this involves the use of much greater computer resources than we have at our disposal at present.

In summary, we have measured the FMR response of arrays of circular ferromagnetic particles.  In addition to the
uniform mode of precession, the FMR spectra shows resonances corresponding to non-uniform spin-wave modes of
precession at values of magnetic field both above and below the peak corresponding to the uniform mode, some of
which are strongly influenced by dipolar interactions between the particles.  Our experimental results and
numerical simulations show that these modes correspond to coupled exchange and dipolar spin-waves.

We thank Anupam Garg for useful discussions, and Robert Tilden for use of computer facilities.  This work was
supported by the Army Research Office through DAAD19-99-1-0339, and by the David and Lucile Packard Foundation.

\end{multicols}

\end{document}